\documentclass[a4paper,12pt]{article}
\usepackage[T2A]{fontenc}
\usepackage[utf8]{inputenc}
\usepackage{vmargin}
\usepackage{amsthm}
\usepackage{amssymb,amsfonts,amsmath,mathtext}
\usepackage{cite,enumerate,float,indentfirst}
\usepackage{caption}
\usepackage[pdftex]{graphicx}

\begin{document}

\title{Non-invariant solutions of the three--dimensional semi--empirical model of the far turbulent wake}
\date{}
\author{O.V. Kaptsov and A.V. Schmidt \\
\\
Institute of Computational Modelling SB RAS, \\
Akademgorodok, Krasnoyarsk, 660036, Russia. \\
e-mail: {\tt kaptsov@icm.krasn.ru} }

\maketitle
\begin{abstract}

A semi-empirical three-dimensional model of turbulence in the approximation of the far turbulent wake behind a body of revolution in a passive stratified medium is considered. The sought quantities are the kinetic turbulent energy, kinetic energy dissipation rate, averaged density defect and density fluctuation variance. The full group of transformations admitted by this model is found. The model is reduced to the system of the ordinary differential equations due to similarity presentations obtained and B--determining equations method. System of ordinary differential equations satisfying natural boundary conditions was solved numerically. The solutions obtained agree with experimental data.
\end{abstract}

\section{Introduction}

The turbulence play an important role in the formation of the ocean structure \cite{Turner, Monin}. For example, the role of turbulence on the evolution of the spatial structure of a thin phytoplankton layer was examined in \cite{Phytoplankton}.

Semi--empirical models of turbulence are now widely used in methods of calculation turbulent flows. However practically there are few analytical approaches to research of this models.

One of the examples of a three--dimensional free turbulent flow is a turbulent wake behind a body of revolution in a stratified medium. Sufficiently complete experimental data on the dynamics of a turbulent wake behind a body of revolution in a linearly stratified medium were obtained by Lin and Pao and presented in \cite{LP}.

The turbulent wake behind an axisymmetrical body in a linearly stratified medium was numerically simulated in \cite{Hassid}. Based on hierarchy of semi--empirical turbulence models of second order, the numerical simulation of the dynamics of a turbulent wake in a stable stratified medium was carried out by Chernykh et al. \cite{Ch1}. A satisfactory agreement with experimental data \cite{LP} was obtained in \cite{Hassid, Ch1}.

A series of papers \cite{TurbSymm1, Shanko, TurbSymm2} was devoted to construction of similarity solutions of semi--empirical turbulence models.
The present paper is a continuation of our investigations. In this paper we consider three--dimensional semi--empirical model of the far turbulent wake behind an axisymmetric self--propelled body in a passive stratified medium \cite{Ch1,Ch2,Ch3}. Considering problem is equivalent to the problem of the development of a turbulent mixing zone in a passive stratified medium \cite{Ch2, Ch3}.

In section 3 we have to define the admissible differential operators of the point groups of transformations \cite{LV,NH} for considering model,
which will allow us to pass to the system of the degenerate elliptic equations. In section 4 we focus on the solutions of the second order to the B--determining equation \cite{BDE} for degenerate elliptic equations. This gives the corresponding differential constraints and allow us
to pass to the system of ordinary differential equations. In section 4 we will present the calculation results.

\section{Model}

To calculate the characteristics of the far turbulent wake behind an axisymmetric self--propelled body in a passive stratified medium we use the three--dimensional semi--empirical turbulence model \cite{Ch1,Ch2,Ch3}
\begin{eqnarray}
\label{model1}
u_0\frac{\partial e}{\partial x}&=&\frac{\partial}{\partial y}C_e\frac{e^2}{\epsilon}\frac{\partial e}{\partial y}+\frac{\partial}{\partial z}C_e\frac{e^2}{\epsilon}\frac{\partial e}{\partial z}-\epsilon,\\
\label{model2}
u_0\frac{\partial \epsilon}{\partial x}&=&\frac{\partial}{\partial y}C_\epsilon \frac{e^2}{\epsilon}\frac{\partial \epsilon}{\partial y}+\frac{\partial}{\partial z}C_\epsilon \frac{e^2}{\epsilon}\frac{\partial \epsilon}{\partial z}-{C_\epsilon}_2\frac{\epsilon^2}{e},\\
\label{model3}
u_0\frac{\partial \langle\rho_1\rangle}{\partial x}&=&\frac{\partial}{\partial y}C_\rho\frac{e^2}{\epsilon}\frac{\partial \langle\rho_1\rangle}{\partial y}+\frac{\partial}{\partial z}C_\rho\frac{e^2}{\epsilon}\frac{\partial \langle\rho_1\rangle}{\partial z}-\frac{\partial }{\partial z}C_\rho\frac{e^2}{\epsilon},\\
\label{model4}
u_0\frac{\partial \langle\rho'^2\rangle}{\partial x}&=&\frac{\partial}{\partial y}{C_1}_\rho\frac{e^2}{\epsilon}\frac{\partial \langle\rho'^2\rangle}{\partial y}+\frac{\partial}{\partial z}{C_1}_\rho\frac{e^2}{\epsilon}\frac{\partial \langle\rho'^2\rangle}{\partial z}+\nonumber\\
&&2C_\rho\frac{e^2}{\epsilon}{\frac{\partial \langle\rho_1\rangle}{\partial y}}^2+2C_\rho\frac{e^2}{\epsilon}\left(\frac{\partial \langle\rho_1\rangle}{\partial z}-1\right)^2-C_T\frac{\langle\rho'^2\rangle\epsilon}{e}.
\end{eqnarray}
In this equations $u_0$ is the velocity of an incoming undisturbed flow, $e(x,y,z)$ is the turbulent kinetic energy, $\epsilon(x,y,z)$ is the kinetic energy dissipation rate, $\langle\rho_1\rangle(x,y,z)$ is the averaged density defect, and $\langle\rho'^2\rangle(x,y,z)$ is the density fluctuation variance. The quantities $C_e=0.136$, $C_\epsilon=0.105$, ${C_\epsilon}_2=1.92$, $C_\rho=0.208$, ${C_1}_\rho=0.087$, $C_T=1.25$ are generally accepted empirical constants \cite{GL,Rodi}.

In what follows, we assume that the velocity of an incoming undisturbed flow equals unity. The marching variable $x$ in equations (\ref{model1})--(\ref{model4}) acts as the time.

By analogy with \cite{TurbSymm1,TurbSymm2}, for the model \eqref{model1}--\eqref{model4} we have to define the admissible differential operators of the point groups of transformations.

\section{Similarity solutions}

Group analysis of the system \eqref{model1}--\eqref{model4} performed by a standard scheme \cite{LV,NH}. The infinitesimal symmetry group of the model (\ref{model1})--(\ref{model4}) is spanned by eight vector fields
\begin{gather}
X_1=\frac{\partial}{\partial x},\quad X_2=\frac{\partial}{\partial y},\quad X_3=\frac{\partial}{\partial z},\quad X_4=\frac{\partial}{\partial \langle\rho_1\rangle},\quad X_5=-z\frac{\partial}{\partial y}+y\frac{\partial}{\partial z}+y\frac{\partial}{\partial \langle\rho_1\rangle}, \nonumber\\
X_6=y\frac{\partial}{\partial y}+z\frac{\partial}{\partial z}+2e\frac{\partial}{\partial e}+2\epsilon\frac{\partial}{\partial \epsilon}+\langle\rho_1\rangle\frac{\partial}{\partial \langle\rho_1\rangle}+2\langle\rho'^2\rangle\frac{\partial}{\partial \langle\rho'^2\rangle},\nonumber\\
X_7=x\frac{\partial}{\partial x}-2e\frac{\partial}{\partial e}-3\epsilon\frac{\partial}{\partial \epsilon},\quad X_8=(\langle\rho_1\rangle-z)\frac{\partial}{\partial \langle\rho_1\rangle}+2\langle\rho'^2\rangle\frac{\partial}{\partial \langle\rho'^2\rangle}.\nonumber
\end{gather}

Next consider the linear combination of the scaling vector fields $X_6$ and $X_7$ 
\begin{equation}
Z=x\frac{\partial}{\partial x}+\alpha y\frac{\partial}{\partial y}+\alpha z\frac{\partial}{\partial z}+2(\alpha-1)e\frac{\partial}{\partial e}+(2\alpha-3)\epsilon\frac{\partial}{\partial \epsilon}+\alpha\langle\rho_1\rangle\frac{\partial}{\partial\langle\rho_1\rangle}+2\alpha\langle\rho'^2\rangle\frac{\partial}{\partial \langle\rho'^2\rangle}.\nonumber
\end{equation}
The solution of the model \eqref{model1}--\eqref{model4} invariant with respect to operator $Z$ has the form
\begin{equation}
e=x^{2\alpha-2}E(\xi,\eta),\quad \epsilon=x^{2\alpha-3}G(\xi,\eta),\quad \langle\rho_1\rangle=x^\alpha H(\xi,\eta),\quad \langle\rho'^2\rangle=x^{2\alpha} R(\xi,\eta),\label{Presentation}
\end{equation}
where $\xi=y/x^\alpha$, $\eta=z/x^\alpha$ is the similarity variables.
Substituting presentation \eqref{Presentation} into \eqref{model1}--\eqref{model4}, we obtain the reduced system
\begin{eqnarray}
C_e\frac{E^2}{G}\left(\frac{\partial^2E}{\partial \xi^2}+\frac{\partial^2E}{\partial \eta^2}\right)-C_e\frac{E^2}{G^2}\left(\frac{\partial E}{\partial \xi}\frac{\partial G}{\partial \xi}+\frac{\partial E}{\partial \eta}\frac{\partial G}{\partial \eta} \right)+2C_e\frac{E}{G}\left(\frac{\partial E}{\partial \xi}^2+\frac{\partial E}{\partial \eta}^2 \right)+\nonumber\\
\alpha\left(\xi\frac{\partial E}{\partial \xi}+\eta\frac{\partial E}{\partial \eta} \right)+2(1-\alpha)E-G=0,\label{ReducedSystem1}\\
C_{\epsilon}\frac{E^2}{G}\left(\frac{\partial^2G}{\partial \xi^2}+\frac{\partial^2G}{\partial \eta^2}\right)+2C_{\epsilon}\frac{E}{G}\left(\frac{\partial E}{\partial \xi}\frac{\partial G}{\partial \xi}+\frac{\partial E}{\partial \eta}\frac{\partial G}{\partial \eta} \right)-C_{\epsilon}\frac{E^2}{G^2}\left(\frac{\partial G}{\partial \xi}^2+\frac{\partial G}{\partial \eta}^2 \right)+\nonumber\\
\alpha\left(\xi\frac{\partial G}{\partial \xi}+\eta\frac{\partial E}{\partial \eta} \right)+(3-2\alpha)G-{C_\epsilon}_2\frac{G^2}{E}=0,\label{ReducedSystem2}\\
C_\rho\frac{E^2}{G}\left(\frac{\partial^2H}{\partial \xi^2}+\frac{\partial^2H}{\partial \eta^2}\right)+2C_\rho\frac{E}{G}\left(\frac{\partial H}{\partial \xi}\frac{\partial E}{\partial \xi}+\frac{\partial H}{\partial \eta}\frac{\partial E}{\partial \eta} \right)+\alpha\left(\xi\frac{\partial H}{\partial \xi}+\eta\frac{\partial H}{\partial \eta} \right)-\nonumber\\
C_\rho\frac{E^2}{G^2}\left(\frac{\partial H}{\partial \xi}\frac{\partial G}{\partial \xi}+\frac{\partial H}{\partial \eta}\frac{\partial G}{\partial \eta} \right)-2C_\rho\frac{E}{G}\frac{\partial E}{\partial \eta}+C_\rho\frac{E^2}{G^2}\frac{\partial G}{\partial \eta}-\alpha H=0,\label{ReducedSystem3}\\
{{C_1}_\rho}\frac{E^2}{G}\left(\frac{\partial^2 R}{\partial \xi^2}+\frac{\partial^2R}{\partial \eta^2}\right)+2{{C_1}_\rho}\frac{E}{G}\left(\frac{\partial R}{\partial \xi}\frac{\partial E}{\partial \xi}+\frac{\partial R}{\partial \eta}\frac{\partial E}{\partial \eta} \right)+\alpha\left(\xi\frac{\partial R}{\partial \xi}+\eta\frac{\partial R}{\partial \eta} \right)-\nonumber\\
{{C_1}_\rho}\frac{E^2}{G^2}\left(\frac{\partial R}{\partial \xi}\frac{\partial G}{\partial \xi}+\frac{\partial R}{\partial \eta}\frac{\partial G}{\partial \eta} \right)+2C_\rho\frac{E^2}{G}\frac{\partial H}{\partial \xi}^2+ 2C_\rho\frac{E^2}{G}\frac{\partial H}{\partial \eta}\left(\frac{\partial H}{\partial \eta}-2\right)+\nonumber\\
2C_\rho\frac{E^2}{G}-C_T\frac{GR}{E}-2\alpha R=0.\label{ReducedSystem4}
\end{eqnarray}

Numerical analysis \cite{Ch1,Ch2,Ch3} of degeneration of the far turbulent wake in a passive stratified medium show that the functions $E$ and $G$ must be presented in the form
\begin{equation}
E(\xi,\eta)=E(\sqrt{\xi^2+\eta^2}),\quad G(\xi,\eta)=G(\sqrt{\xi^2+\eta^2}).\label{SimilarityDeg}
\end{equation}
Note that the presentations \eqref{SimilarityDeg} are satisfied all the reduced equations \eqref{ReducedSystem1}--\eqref{ReducedSystem4}. Changing to polar coordinates $\xi=r\cos(\phi)$, $\eta=r\sin(\phi)$, and by virtue of \eqref{SimilarityDeg} the reduced system become
\begin{eqnarray}
C_e\frac{E}{G}\left(EE''+2E'^2-\frac{E}{G}E'G'+\frac{E}{r}E'\right)+\alpha rE'+2(1-\alpha)E-G=0,\label{Polar1}\\
C_{\epsilon}\frac{E}{G}\left(EG''-\frac{E}{G}G'^2+2E'G'+\frac{E}{r}G'\right)+\alpha rG'+(3-2\alpha)G-{C_{\epsilon}}_2\frac{G^2}{E}=0,\label{Polar2}\\
C_\rho\frac{E^2}{G}\left(H_{rr}+\frac{1}{r^2}H_{\phi\phi}\right)+\left(C_\rho\frac{E}{G}\left(2E'-\frac{E}{G}G'+\frac{E}{r} \right)+\alpha r\right)H_r-\alpha H+\nonumber\\
C_\rho\frac{E}{G}\left(\frac{E}{G}G'-2E'\right)\sin(\phi)=0,\label{Polar3}\\
{{C_1}_\rho}\frac{E^2}{G}\left(R_{rr}+\frac{1}{r^2}R_{\phi\phi}\right)+\left({{C_1}_\rho}\frac{E}{G}\left(\frac{E}{r}+2E'-\frac{E}{G}G'\right)+\alpha r\right)R_r-\nonumber\\
\left(C_T\frac{E}{G}+2\alpha\right)R+2C_\rho\frac{E^2}{G}\left(H_{r}^2+\frac{1}{r^2}H_{\phi}^2\right)+2C_\rho\frac{E^2}{G}-\nonumber\\
4C_\rho\frac{E^2}{G}\left(\sin(\phi)H_r+\frac{\cos(\phi)}{r}H_\phi\right)=0,\label{Polar4}
\end{eqnarray}
where $E=E(r)$, $G=G(r)$, $H=H(r,\phi)$, $R=R(r,\phi)$. Here and elsewhere, subscripts denote derivatives, so $H_r=\partial H/ \partial r$, etc. We now apply the BDE method \cite{BDE} to reduce the equations \eqref{Polar3}, \eqref{Polar4} to some ordinary differential equations.

\section{BDE method}

Consider more general equation than \eqref{Polar3}
\begin{equation}
H_{\phi\phi}+r^2H_{rr}+A(r)H_r+B(r)H+C(r)\sin(\phi)=0,\label{PDE1}
\end{equation}
where $A(r)$, $B(r)$, $C(r)$ are arbitrary functions. We take the B--determining equation corresponding to \eqref{PDE1} of the form 
\begin{equation}
D_\phi^2 h+r^2D_r^2h+b_1(r,\phi)D_rh+b_2(r,\phi)h=0.\label{BDequation}
\end{equation}
Here and throughout $D_\phi, D_r$ are the operators of total differentiation with respect to $\phi$ and $r$. The functions $b_1(r,\phi)$ and $b_2(r,\phi)$ are to be determined together with the function $h$. Note that for classical determining equations \cite{LV,NH} holds
\begin{equation}
b_1(r,\phi)=A(r), \quad b_2(r,\phi)=B(r).\nonumber
\end{equation}
We seek second order solution of \eqref{BDequation} of the form
\begin{equation}
h=H_{\phi\phi}+h_1\left(\phi,H,H_\phi\right).\label{BDE Solution}
\end{equation}
Substituting \eqref{BDE Solution} into BDE \eqref{BDequation} leads to an equation which includes derivatives of the fourth order. We can express the derivatives $H_{rr\phi\phi}$, $H_{\phi\phi\phi\phi}$, $H_{r\phi\phi}$, $H_{\phi\phi\phi}$, $H_{\phi\phi}$ by means of \eqref{PDE1}. Setting the coefficient of $H_{rrr}$ equal to zero we obtain $b_1(r,\phi)=A(r)$.

The left-hand side of \eqref{BDequation} is a polynomial with respect to $H_{rr}$ and $H_{r\phi}$. This polynomial must identically vanish. Collecting similar terms we obtain the equations
\begin{equation}
\begin{split} 			\label{BDE Sys 1}
h_{1_{H_\phi H_\phi}}=0, \quad h_{1_{H H_\phi}}=0, \quad 2\left(A(r)H_r+B(r)H+C(r)\sin(\phi) \right)h_{1_{H_\phi H_\phi}}-\\
2H_\phi h_{1_{H H_\phi}}-2h_{1_{\phi H_\phi}}+B(r)-b_2(r,\phi)=0.
\end{split}
\end{equation}
It is easy to show that the general solution of the equations \eqref{BDE Sys 1} is
\begin{equation}
h_1\left(\phi,H,H_\phi\right)=h_2(\phi)H_\phi+h_3(\phi,H), \quad b_2(r,\phi)=B(r)-2h_2'(\phi).\nonumber
\end{equation}

Substituting the functions $b_1$, $b_2$ and $h_1$ into BDE \eqref{BDequation} we obtain that the left-hand side of \eqref{BDequation} is a polynomial with respect to $H_{r}$ and $H_\phi$. This polynomial must identically vanish. Collecting similar terms leads to the following equations
\begin{equation}
\begin{split}  \label{BDE Sys 2}
h_{3_{HH}}=0, \quad 2h_{3_{\phi H}}+h_2''(\phi)-2h_2'(\phi)h_2(\phi)=0, \quad (B(r)H+C(r)\sin(\phi))h_{3_H}-\\
h_{3_{\phi\phi}}+(2h_2'(\phi)-B(r))h_3+C(r)(\cos(\phi)h_2(\phi)-\sin(\phi))=0.
\end{split}
\end{equation}
The equations \eqref{BDE Sys 2} imply
\begin{gather}
h_3(\phi,H)=\left(\frac 1 2h_2(\phi)^2-\frac 1 2 h_2'(\phi)+h_4\right)H,\nonumber\\
h_2'(\phi)-h_2^2(\phi)-2\cot(\phi)h_2(\phi)+2(1-h_4)=0,\label{Riccati Eq}
\end{gather}
here $h_4$ is arbitrary constant.

Clearly, that the Riccati equation \eqref{Riccati Eq} has the partial solution
\begin{equation}
h_2(\phi)=\tan(\phi)
\end{equation}
for $h_4=1/2$.

Thus we find the second order solution of the BDE \eqref{BDequation}
\begin{equation}
h=H_{\phi\phi}+\tan(\phi)H_\phi.\nonumber
\end{equation}
The corresponding differential constraint $h=0$ has the general solution
\begin{equation}
H=H_1(r)\sin(\phi)+H_2(r),\label{Density defect}
\end{equation}
where $H_1$ and $H_2$ are arbitrary functions.

Substitution \eqref{Density defect} into equation \eqref{Polar4} gives
\begin{eqnarray}
{{C_1}_\rho}\frac{E^2}{G}\left(R_{rr}+\frac{1}{r^2}R_{\phi\phi}\right)+\left({{C_1}_\rho}\frac{E}{G}\left(\frac{E}{r}+2E'-\frac{E}{G}G'\right)+\alpha r\right)R_r-\nonumber\\
\left(C_T\frac{E}{G}+2\alpha\right)R+2C_\rho\frac{E^2}{r^2G}\Bigl((rH_1'-H_1)(rH_1'+H_1-2r)\sin^2(\phi)+\nonumber\\
2r^2(H_1'-1)H_2'\sin(\phi)+r^2H_2'^2+(H_1-r)^2\Bigr)=0.\label{Polar4 subs H}
\end{eqnarray}

By analogy with the case of the equation \eqref{Polar3}, consider more general equation than \eqref{Polar4 subs H}
\begin{equation}
R_{\phi\phi}+r^2R_{rr}+K(r)R_r+L(r)R+M(r)\sin^2(\phi)+N(r)\sin(\phi)+P(r)=0,\label{Polar4 General}
\end{equation}
where $K(r)$, $L(r)$, $M(r)$, $N(r)$ and $P(r)$ are arbitrary functions. The BDE method applied to equation \eqref{Polar4 General} gives rise to the following results:
\begin{gather}
b_1(r,\phi)=K(r), \quad b_2(r,\phi)=L(r)-\frac{8}{\sin^2(2\phi)},\nonumber\\
h=R_{\phi\phi}-2\cot(2\phi)R_{\phi},\label{BDE Solution for Polar4}\\
N(r)=0.\label{BDE case}
\end{gather}

The formula \eqref{BDE case} for the equation \eqref{Polar4 subs H} takes the form
\begin{equation}
(H_1'-1)H_2'=0.\nonumber
\end{equation}
Clearly, that we must explain the case
\begin{equation}
H_2'=0.\label{Presentation case}
\end{equation}
Integrating differential constraint $h=0$ corresponding to the BDE solution \eqref{BDE Solution for Polar4}, we find
\begin{equation}
R=R_1(r)\sin^2(\phi)+R_2(r),\label{Density fluctuation variance}
\end{equation}
where $R_1(r)$ and $R_2(r)$ are arbitrary functions.

Thus in the similarity variables $\xi$ and $\eta$ from \eqref{Density defect}, \eqref{Presentation case}, \eqref{Density fluctuation variance} we have
\begin{equation}
H(\xi,\eta)=H_3(\sqrt{\xi^2+\eta^2})\eta + H_2, \quad R(\xi,\eta)=R_3(\sqrt{\xi^2+\eta^2})\eta^2 + R_2(\sqrt{\xi^2+\eta^2}),\label{SimilarityDeg2}
\end{equation}
where $H_3=H_1/\sqrt{\xi^2+\eta^2}$ and $R_3=R_1/(\xi^2+\eta^2)$.

This allow us to reduce the model \eqref{model1}--\eqref{model4} to the system of ordinary differential equations. Substituting presentations \eqref{SimilarityDeg}, \eqref{SimilarityDeg2} into the reduced system \eqref{ReducedSystem1}--\eqref{ReducedSystem4} we obtain
\begin{eqnarray}
&&H_2=0,\nonumber\\
&&E''=E'\left(\frac{G'}{G}-2\frac{E'}{E}-\frac{1}{\tau}\right)+\frac{G}{C_eE}\left(2(\alpha-1)+\frac{G}{E}-\alpha\tau\frac{E'}{E}\right),\label{ODE 1}\\
&&G''=G'\left(\frac{G'}{G}-2\frac{E'}{E}-\frac{1}{\tau}\right)+\frac{G}{C_{\epsilon}E^2}\left((2\alpha-3)G+\frac{{C_{\epsilon}}_2G^2}{E}-\alpha\tau G'\right),\label{ODE 2}\\
&&H_3''=H_3'\left(\frac{G'}{G}-2\frac{E'}{E}-\frac{3}{\tau}-\frac{\alpha\tau G}{C_\rho E^2}\right)+\frac{H_3-1}{\tau}\left(\frac{G'}{G}-2\frac{E'}{E}\right),\label{ODE 3}\\
&&R_3''=R_3'\left(\frac{G'}{G}-2\frac{E'}{E}-\frac{5}{\tau}-\frac{\alpha\tau G}{{C_1}_\rho E^2}\right)+\frac{2R_3}{\tau}\left(\frac{G'}{G}-2\frac{E'}{E}+\frac{C_T\tau G^2}{2{C_1}_\rho E^3}\right)-\nonumber\\
&&\frac{2C_\rho H_3'}{{C_1}_\rho}\left(\frac{2(H_3-1)}{\tau}+H_3'\right),\label{ODE 4}\\
&&R_2''=R_2'\left(\frac{G'}{G}-2\frac{E'}{E}-\frac{1}{\tau}-\frac{\alpha\tau G}{{C_1}_\rho E^2}\right)+\frac{R_2G}{{C_1}_\rho E^2}\left(\frac{C_TG}{E}+2\alpha\right)-\nonumber\\
&&2R_3-\frac{2C_\rho}{{C_1}_\rho}(H_3-1)^2,\label{ODE 5}
\end{eqnarray}
where $\tau=\sqrt{\xi^2+\eta^2}$.

\section{Calculation results}

System \eqref{ODE 1}--\eqref{ODE 5} has to satisfy the conditions
\begin{eqnarray}
E'=G'=H_1'=R_1'=R_2'=0, \tau=0, \label{Cond Symm}\\
E=G=H_1=R_1=R_2=0, \tau\to\infty. \label{Cond Und}
\end{eqnarray}

Conditions \eqref{Cond Symm} takes into account flow symmetry with respect to the OX axis. The boundary conditions \eqref{Cond Und} imply
that all functions take zero values outside the turbulent wake.

\begin{figure}[H]
\begin{minipage}[h]{0.47\linewidth}
\center{\includegraphics[width=1\linewidth]{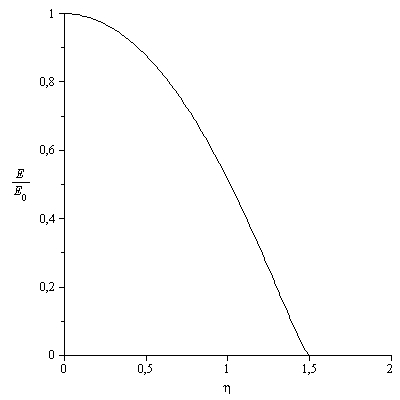}} (a) \\
\end{minipage}
\hfill
\begin{minipage}[h]{0.47\linewidth}
\center{\includegraphics[width=1\linewidth]{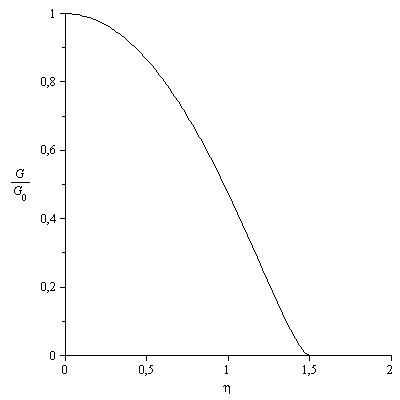}} (b) \\
\end{minipage}
\vfill
\vskip+1\baselineskip
\begin{minipage}[h]{0.47\linewidth}
\center{\includegraphics[width=1\linewidth]{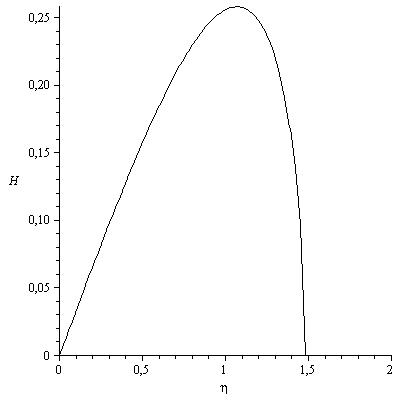}} (c) \\
\end{minipage}
\hfill
\begin{minipage}[h]{0.47\linewidth}
\center{\includegraphics[width=1\linewidth]{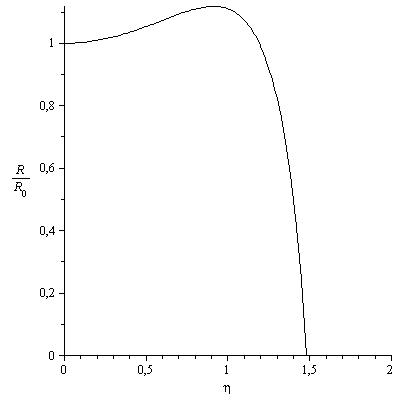}} (d) \\
\end{minipage}
\caption{Calculated profiles as $\xi=0$: (a) normed profile of $E$, (b) normed profile of $G$, (c) profile of $H$, (d) normed profile of $R$.}
\end{figure}

The system \eqref{ODE 1}--\eqref{ODE 5} of ordinary differential equations satisfying boundary condition \eqref{Cond Symm}, \eqref{Cond Und} was solved numerically. Additional difficulties are caused by the fact that the coefficients of ordinary differential equations have singularities. The problem was solved by a modified shooting method and asymptotical expansion of the solution in the vicinity of the singular point \cite{Shanko}.

\begin{figure}[H]
\begin{minipage}[h]{0.47\linewidth}
\center{\includegraphics[height=70mm]{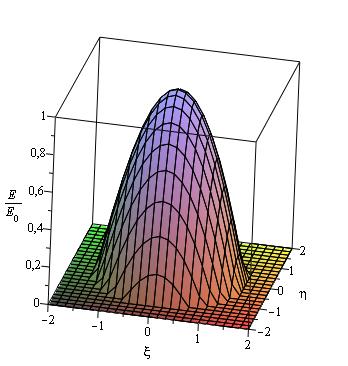}} (a) \\
\end{minipage}
\hfill
\begin{minipage}[h]{0.47\linewidth}
\center{\includegraphics[height=70mm]{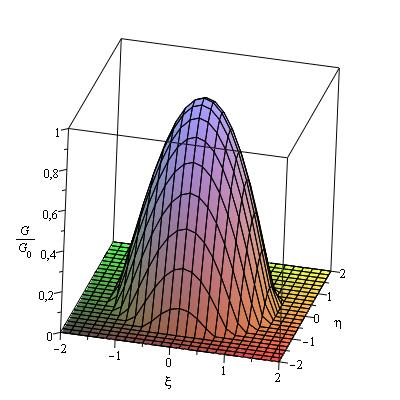}} (b) \\
\end{minipage}
\vfill
\begin{minipage}[h]{0.47\linewidth}
\center{\includegraphics[height=70mm]{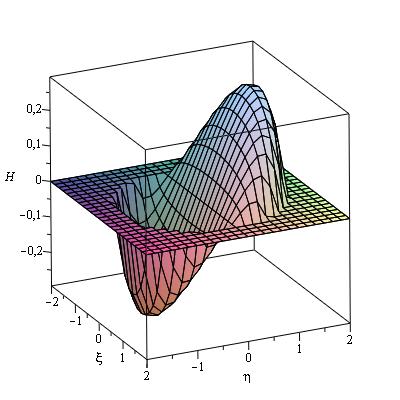}} (c) \\
\end{minipage}
\hfill
\begin{minipage}[h]{0.47\linewidth}
\center{\includegraphics[height=70mm]{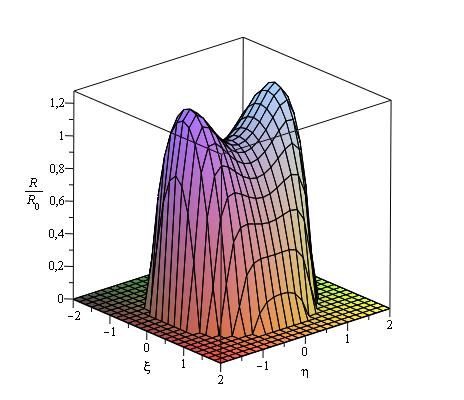}} (d) \\
\end{minipage}
\caption{Calculated functions: (a) function $E/E_0$, (b) function $G/G_0$, (c) function $H$, (d) function $R/R_0$.}
\end{figure}

Value of $\alpha$ a taken to be equal to $0.23$. The results for the problem solution are illustrated in Figs. 1, 2. Figure 1 shows the profiles of the functions $E/E_0, G/G_0, H$ and $R/R_0$ as $\xi=0$, where subscript $0$ denote axial value. The functions $E/E_0, G/G_0, H$ and $R/R_0$ are plotted in Fig. 2.

The function $H(0,\eta)$ characterizing the degree of fluid mixing in the turbulent wake a given in Fig. 1c. As can be seen, the maximum value of this function slightly differ from $0.25$, which is consistent with the present notions of incomplete fluid mixing in the wakes \cite{TurbMix}.

In Fig. 3 the axial values of the turbulent energy are compared with Lin and Pao`s experimental data \cite{LP}, Hassid`s computational results \cite{Hassid} and results of numerical computations \cite{Ch1, Ch2}. We have borrowed this figure from work \cite{Ch1} and have put the values. We can see satisfactory agreement with Lin and Pao`s experimental data here as well.

\begin{figure}
\center{\includegraphics[width=0.47\linewidth]{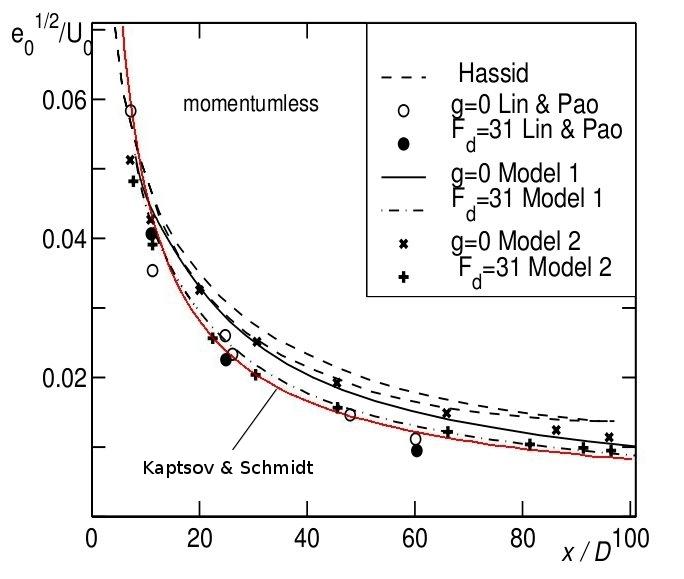}}
\caption{Axial values of the turbulent energy.}
\end{figure}

\section*{Conclusion}

The main results of the paper are as follows. The three--dimensional semi--empirical turbulence model of the far turbulent wake behind an axisymmetric self--propelled body in a passive stratified medium was reduced to the system of ordinary differential equations due to similarity presentations obtained and B--determining equations method. The system of ordinary differential equations satisfying natural boundary conditions was solved numerically. The solutions constructed agree with experimental data.

\section*{Acknowledgements}

The authors are grateful to G.G. Chernykh for the problem formulation, materials presented, and useful discussions.

This work was supported by the Russian Foundation for Basic Research (grant Nos. 07-01-00489-a and 07-01-00363-a), President's Grant "Leading Scientific Schools" NSh-7256.2010.1 and Siberian Branch of Russian Academia of Science (grant No. 103).

\end{document}